\documentclass{article}

\usepackage{PRIMEarxiv}

\usepackage[utf8]{inputenc} 
\usepackage[T1]{fontenc}    
\usepackage{hyperref}       
\usepackage{url}            
\usepackage{booktabs}       
\usepackage{amsfonts}       
\usepackage{nicefrac}       
\usepackage{microtype}      
\usepackage{lipsum}
\usepackage{fancyhdr}       
\usepackage{graphicx}       
\graphicspath{{media/}}     

\usepackage{bm}
\usepackage{adjustbox}
\usepackage{amsmath}
\usepackage{caption} 
\title{An Overview on Generative AI at Scale with Edge-Cloud Computing}

\author{
  Yun-Cheng Wang \\
  University of Southern California \\
  Los Angeles, USA\\
  \texttt{yunchenw@usc.edu} \\
   \And
  Jintang Xue \\
  University of Southern California \\
  Los Angeles, USA\\
  \texttt{jintangx@usc.edu} \\
   \And
  Chengwei Wei \\
  University of Southern California \\
  Los Angeles, USA\\
  \texttt{chengwei@usc.edu} \\
   \And
  C.-C. Jay Kuo \\
  University of Southern California \\
  Los Angeles, USA\\
  \texttt{cckuo@sipi.usc.edu} \\
}

\begin{document}
\maketitle

\begin{abstract}
As a specific category of artificial intelligence (AI), generative
artificial intelligence (GenAI) generates new content that resembles
what is created by humans. The rapid development of GenAI systems has
created a huge amount of new data on the Internet, posing new challenges
to current computing and communication frameworks. Currently, GenAI
services rely on the traditional cloud computing framework due to the
need for large computation resources. However, such services will
encounter high latency because of data transmission and a high volume of
requests. On the other hand, edge-cloud computing can provide adequate
computation power and low latency at the same time through the
collaboration between edges and the cloud. Thus, it is attractive to build
GenAI systems at scale by leveraging the edge-cloud computing
paradigm.  In this overview paper, we review recent developments in
GenAI and edge-cloud computing, respectively.  Then, we use two
exemplary GenAI applications to discuss technical challenges in scaling
up their solutions using edge-cloud collaborative systems. Finally, we
list design considerations for training and deploying GenAI
systems at scale and point out future research directions. 
\end{abstract}


\section{Introduction}\label{sec:introduction}

Generative AI (GenAI) has emerged as a groundbreaking
field to realize artificial general intelligence (AGI) by integrating
machine learning and creative content generation. It is a specific
category of AI that aims to autonomously generate new content that
imitates the content created by humans in different modalities,
including images \cite{goodfellow2020generative, rombach2022high}, audio
\cite{ren2019fastspeech, ren2020fastspeech}, 
text \cite{brown2020language, touvron2023llama}, and even 3D
objects \cite{martin2022scangan360, nichol2022point}. 
With the rapid development of
GenAI, various applications, such as text-to-image generation
\cite{ramesh2022hierarchical}, text-to-speech (TTS) synthesis
\cite{tan2022naturalspeech}, 
chatbots \cite{openai2023gpt4}, and AI-rendered virtual reality (VR)
\cite{ratican2023proposed}, have been proposed and are
widely used by consumers. However, since most GenAI models have huge
model sizes and are computationally demanding, a powerful centralized
computation infrastructure (i.e. cloud server) is required to process
requests from users. As a result, users may experience high latency if
traffic volumes are high. Such limitations hinder their applicability to
high-volume applications that require low latency.  Besides, the heavy
computation in a cloud consumes a significant amount of energy. The
overly-centralized computing framework is eco-unfriendly, unsustainable,
and cost-inefficient. 

In recent years, the proliferation of mobile devices and the exponential
growth of data-intensive applications have spurred the development of
edge-cloud computing solutions. Edge-cloud computing takes advantage of
powerful computation resources in cloud servers and efficient data
management and communication in edge servers. It has emerged as a
promising solution for consumer-based AI applications and edge
intelligence.  For example, several large AI models are deployed with
the edge-cloud computing system \cite{qi2018enabling, xiao2021towards}.
Compared to traditional cloud computing, which focuses on computation in
cloud servers, and multi-access edge computing (MEC), which focuses on
computation in edge servers, edge-cloud computing can exploit more
computation resources and lower latency through the collaboration
between clouds and edges. 

\begin{figure}
\centerline{\includegraphics[width=0.6\linewidth]{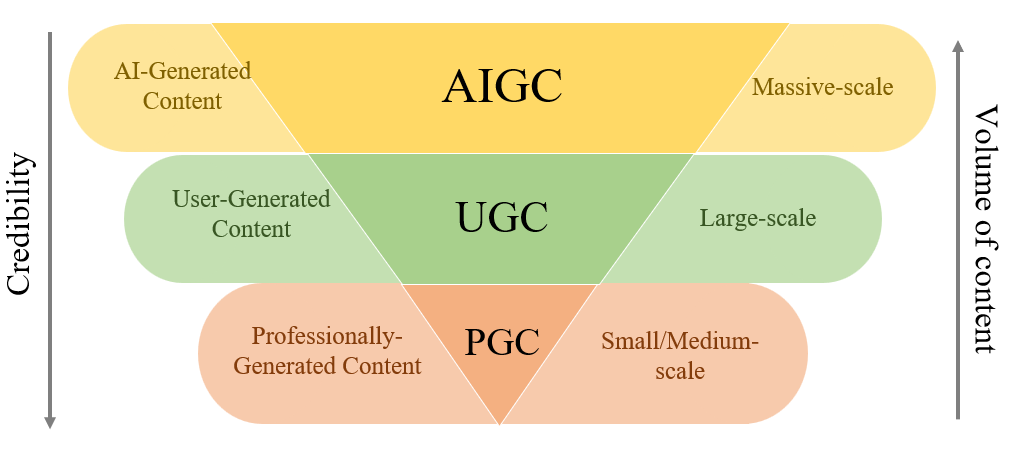}}
\caption{Different phases of content creation.}\label{fig:scale}
\end{figure}

GenAI poses unprecedented challenges to scalable computing systems and
the need for edge-cloud computing because of three reasons. First, the transmission
latency increases significantly in GenAI services due to the tremendous
amount of data generated by AI, or so-called AI-generated content
(AIGC). Fig.  \ref{fig:scale} shows the evolution of
different phases in content creation. Compared to
professionally-generated content (PGC) and user-generated content (UGC),
GenAI created much more data on the Internet. As a result, transmission
latency becomes a serious challenge in GenAI services. 
The second challenge is the deployment of computation systems. Currently, most GenAI
services target consumer-centric applications. It makes more sense to
place the computation system closer to users instead of relying on a
centralized computation infrastructure to process all user requests. In
addition, edge-cloud computing can preserve more privacy for users by
storing their data only on local servers or user devices. 
Third, the required resources to run GenAI services are huge. 
For example, ChatGPT by OpenAI\footnote{\label{note:chatgpt}\url{https://openai.com/blog/chatgpt}} 
is one of the most popular GenAI services recently. It's a chatbot to
answer users' questions in human-like responses interactively.
It processed more than 13 million requests per day in January 2023 
\cite{wu2023ai}. 
Although the exact computing infrastructure used by the ChatGPT service
is not publicly available, we can estimate the 
cost to run the service each day based on the model architecture of
GPT-3 \cite{brown2020language}, the generative model to support
the ChatGPT service.
GPT-3 is a large language model (LLM) containing 175 billion parameters,
which requires more than 350 GB of RAM and VRAM to run the model.
To deploy such a large model, a distributed computing system with 
at least 2,048 CPUs and 2,048 GPUs in order to handle arbitrary
user inputs and achieve reasonable latency. 
Relying solely on the computation power in the cloud will lead to high 
latency when the request volume is high.
In addition, its daily electricity charge is estimated to be 
around \$600,000 using Nvidia A100 GPUs.
Not to mention the training of GPT-3, which requires $10^8$ 
times computation and more than $10^5$ iterations. Thus, it's
unsustainable, and cost-inefficient to deploy such a service
entirely on the cloud servers.
Collaboration of edge and cloud computing resources will mitigate 
the burden of cloud servers.  In this paper, 
we examine four aspects of deploying GenAI with edge-cloud computing: 
1) computation and data offloading, 2) low latency, 3) personalization, and 4) privacy.

\begin{figure}
\centerline{\includegraphics[width=\linewidth]{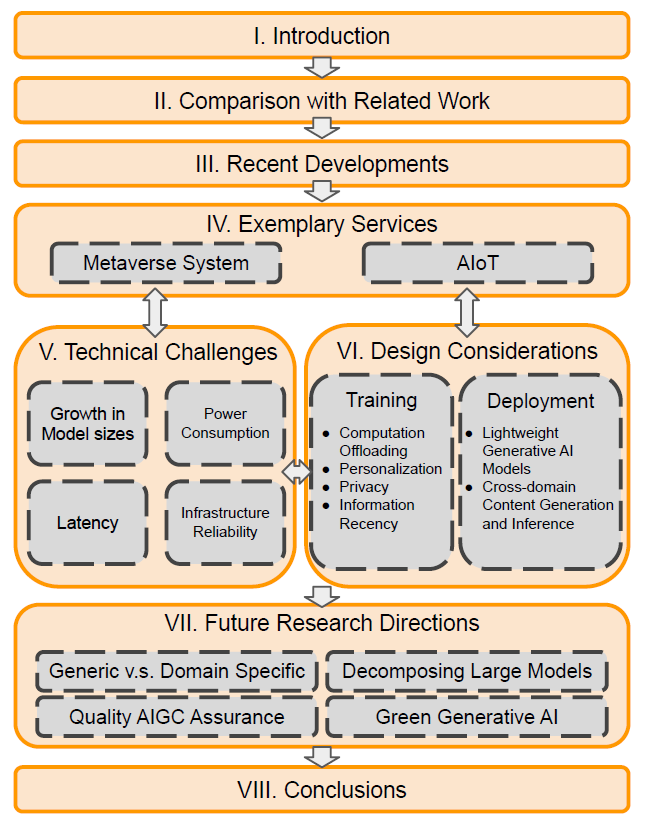}}
\caption{The organization of this paper.}\label{fig:scope}
\end{figure}

The flow of this paper is illustrated in Fig. \ref{fig:scope}.
Our main contributions are summarized below:
\begin{itemize}
\item Provision of a comprehensive overview of recent developments in
both GenAI models and edge-cloud computing;
\item Identification of technical challenges in the deployment of
large-scale GenAI models using today's solution;
\item Presentation of design considerations that target computational
efficiency (i.e. lower power consumption), low latency, personalization,
and privacy;
\item Visualization of two large-scale GenAI applications as concrete examples 
to support our discussion;
\item Future research directions on GenAI systems based on edge-cloud 
computing.
\end{itemize}

The rest of this paper is organized below. Comparison of this work and
related previous overview papers is made in Sec.  \ref{sec:related}.
Reviews on recent developments of GenAI and edge-cloud computing are
conducted in Sec.  \ref{sec:background}. Two application scenarios are
envisioned in Sec.  \ref{sec:app}. Technical challenges in deploying
GenAI systems at scale with current distributed systems are examined in
Sec.  \ref{sec:challenges}. Design considerations to address them with
edge-cloud computing are elaborated in Sec.  \ref{sec:design}. Finally,
future directions are pointed out in Sec.  \ref{sec:future}, and
concluding remarks are given in Sec.  \ref{sec:conclusion}. 

\begin{figure*}[t]
     \centering
  \begin{tabular}{c@{}c@{}c}
    \includegraphics[width=.32\linewidth]{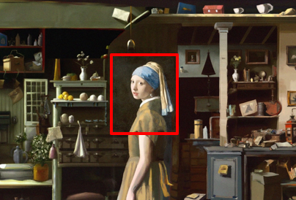} &
    \includegraphics[width=.32\linewidth]{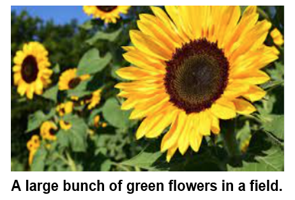} &
    \includegraphics[width=.32\linewidth]{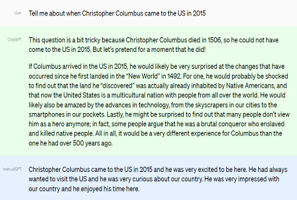}
    \\[\abovecaptionskip]
    \small (a) Image outpainting\footref{note:dalle}. & \small (b) Image captioning. & \small (c) Chatbot\footref{note:chatgpt}.\\
    \includegraphics[width=.32\linewidth]{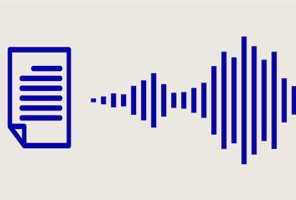} &
    \includegraphics[width=.32\linewidth]{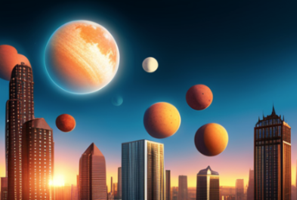} &
    \includegraphics[width=.32\linewidth]{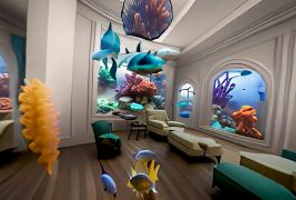}
    \\[\abovecaptionskip]
    \small (d) Text-to-speech (TTS). & \small (e) AI art. & \small (f) Metaverse.
  \end{tabular}
\caption{Six illustrative applications of GenAI models: a) image
outpainting, b) image captioning, c) chatbot, d) text-to-speech, e)
AI art, and f) metaverse.}\label{fig:genai_app}
\end{figure*}

\section{COMPARISON WITH RELATED WORK} \label{sec:related}


We summarize related overview papers and compare them with this work
below.

\subsection{GENERATIVE AI}

After the release of ChatGPT at the end of 2022, interest in GenAI
increase rapidly, and a few survey or overview papers on GenAI have been
published \cite{cao2023comprehensive, wu2023ai, zhang2023complete}.
Some focus on how GenAI models can be applied to different applications,
such as audio diffusion \cite{zhang2023survey}, text-to-image generation
\cite{zhang2023text}, and multimodality \cite{suzuki2022survey}
applications. Nevertheless, most of them are concerned with the
algorithmic aspect of GenAI. Here, we study technical challenges related
to the deployment of GenAI systems at scale and propose a practical
cloud-edge computing solution. 

\subsection{EDGE INTELLIGENCE}

There are plenty of survey and overview papers on AI in edge-cloud
computing, or so-called ``edge intelligence". Most of them consider
discriminative AI tasks, where the systems only need to make binary
decisions, such as edge intelligence in surveillance cameras
\cite{yao2022edge, hua2023edge}, which is an important topic for
security. Other emerging edge intelligence applications include unmanned
autonomous vehicles (UAV) \cite{mcenroe2022survey} and the Internet of
Things (IoT) \cite{firouzi2022convergence}. The latter has long been an
important field since the arrival of 5G and wireless networks.  A
roadmap about the integration of edge-cloud computing and AI is given in
\cite{ding2022roadmap}. 

GenAI has become a new application domain of AI technology in recent years.
It poses emerging challenges, including a huge amount of machine-created
content, large model sizes and power consumption, and low latency
requirements in some real-time applications, such as GenAI for gaming.
It is a critical problem since the amount of transmitted content is much more
than discriminant AI tasks.  To the best of our knowledge,
\cite{xu2023unleashing} is the only work that addressed GenAI at the
edge.  However, it focused on the review of existing papers. In this
work, we not only provide a comprehensive review of recent developments
of GenAI and edge-cloud computing but also have an in-depth discussion
on many related issues, including technical challenges, design
considerations, exemplary applications, and future technology outlook of
GenAI deployment at scale using the edge-cloud platform. 

\section{RECENT DEVELOPMENTS IN GENERATIVE AI AND 
EDGE-CLOUD COMPUTING} \label{sec:background}


\begin{figure}[t]
     \centering
  \begin{tabular}{c}
    \includegraphics[width=0.8\linewidth]{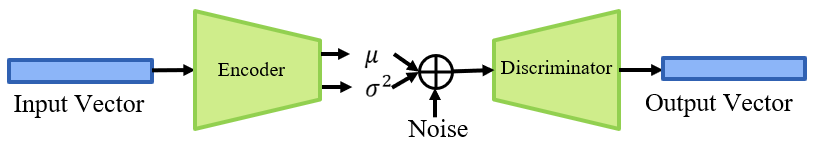}
    \\[\abovecaptionskip]
    \small (a) Variational Auto-encoder (VAE).\\ \\
    \includegraphics[width=0.8\linewidth]{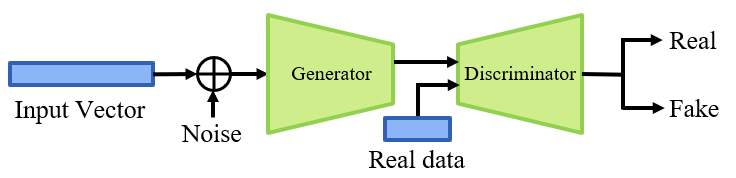}
    \\[\abovecaptionskip]
    \small (b) Generative Adversarial Network (GAN).\\ \\
    \includegraphics[width=0.64\linewidth]{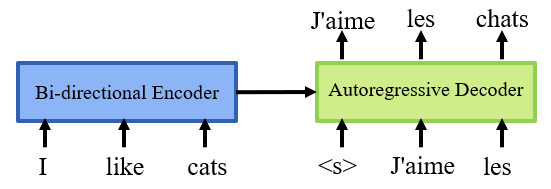}
    \\[\abovecaptionskip]
    \small (c) Transformer.\\ \\
  \end{tabular}
\caption{Architectures of three popular GenAI model categories: 
VAE, GAN, and Transformer.}\label{fig:gen-arch}
\end{figure}



\subsection{GENERATIVE AI}

With the explosion of ChatGPT, GenAI has become a hot topic. GenAI is an
AI technology that can generate various multimedia content
\cite{oussidi2018deep, harshvardhan2020comprehensive, suzuki2022survey}. 
Fig. \ref{fig:genai_app} shows some real-world applications of GenAI,
including images, texts, audio, graphics, and even 3D objects.  The
historical development of generative AI can be roughly divided into
three eras: 1) the Variational Autoencoder (VAE) and Generative 
Adversarial Network (GAN) era (2014-2017), 2) the Transformer
era (2018-2019), and 3) the large model era (2020-present)
\cite{foster2019generative}.  Three popular architectures for GenAI
models are shown in Fig. \ref{fig:gen-arch}. 

The Variational Autoencoder (VAE) was first proposed in
\cite{kingma2013auto}. It has several variations
\cite{wei2020variations, girin2020dynamical, zhai2018autoencoder,
asperti2021survey} to improve the quality of the generated content
\cite{tolstikhin2017wasserstein, larsen2016autoencoding}, 
adjust to different levels of supervision \cite{gao2020zero},
and improve the inference efficiency \cite{sohn2015learning}. 
VAEs are probabilistic generative models. 
Their encoder and decoder correspond to two neural
networks. An encoder maps an input to a vector in a latent space, while
the decoder maps a latent vector back to the input space to generate an
output. In the training stage, the network parameters are optimized so
that the output is as close as possible to the input. Adding noise to
latent vectors makes the decoder produce multiple output samples that
have the same distribution as input samples. 

Similar to VAE, Generative Adversarial Networks (GANs)
\cite{goodfellow2020generative} need two
networks in the training stage but keep only one in the inference stage
\cite{wang2021generative, pan2019recent, hong2019generative, creswell2018generative}.
The two networks are a generator and a discriminator. Through a training
process \cite{gui2021review, jabbar2021survey}, 
the generator generates better and better fake data that are
getting closer to real data in the distribution to fool the
discriminator.  On the other hand, the discriminator is used to
differentiate real and fake data as much as possible. The generator
and discriminator are trained by solving a min-max optimization problem: 
$$
\min_G \max_D V(G, D) = \mathop{\mathbb{E}}_{x\sim \text{real}}[\log D(x)] +  \mathop{\mathbb{E}}_{G(z)\sim \text{fake}}[1 - \log D(G(z))]
$$
where $G(*)$ and $D(*)$ denotes the generator and discriminator, respectively.
Its capability improves along the
training process. Gradually, they reach an equilibrium status where fake
and real data are so close that they cannot be easily differentiated.
Then, the training stage is completed.

Natural language generation (NLG) models aim to generate human-like textual 
responses. There are several common applications, such as 
neural machine translation \cite{cho2020learning, sutskever2014sequence},
question answering \cite{clark2020electra, yang2019xlnet}, and 
document summarization \cite{wang2022focused, nema2017diversity}. 
Such models are also called language models (LMs) \cite{wei2023overview}.
In recent years, transformers \cite{vaswani2017attention} with
self-attention mechanisms have made major breakthroughs in establishing
powerful LMs \cite{han2022survey, khan2022transformers,
tay2022efficient, liu2023survey, selva2023video}. Transformers have replaced the
long short-term memory (LSTM) \cite{hochreiter1997long} as the preferred
LM architecture, and set off a new wave of large language models (LLMs) 
\cite{lin2022survey, kalyan2021ammus, zhang2022survey, du2022survey}.
They often adopt an encoder-decoder architecture as shown in 
Fig. \ref{fig:gen-arch} (c). While the encoder adopts
a bi-directional information propagation process to understand the input 
text, the decoder in most transformer architectures generates words one by
one. Such a decoder is also called the autoregressive decoder.
With the advent of transformers, generative models are getting larger
and larger. Over the past two years, attempts have been made to combine
a wide variety of models to create larger and more powerful models.
They offer impressive performance in various fields \cite{gozalo2023chatgpt}. 
Due to the large model sizes of GenAI models, they are deployed on
the cloud nowadays. That is, models are trained at the training stage
and run at the inference stage in cloud servers.  Users send requests to
the cloud server for content generation.  Then, the generated content is
sent back to users. 

However, due to the long distance between users and the cloud, the
above-mentioned framework is not scalable.  It has higher generation
latency, which hinders specific applications such as augmented reality
(AR) / virtual reality (VR) / mixed reality (MR). 
Furthermore, with the rapid growth of GenAI services, the amount
of AI-generated data on the Internet has increased significantly (see
Fig. \ref{fig:scale}).  Some GenAI web-based services, such as
mid-journey\footnote{\url{https://www.midjourney.com/}} and 
DALL-E\footnote{\label{note:dalle}\url{https://openai.com/blog/dall-e-introducing-outpainting}}, 
have a large number of users per day.  Most
GenAI services are not free since the required computation is costly due
to high power consumption.  Latency can be another major concern once
the service becomes popular with growing user requests. 

It is worthwhile to emphasize that user feedback is important for model
fine-tuning. In other words, there are interactions between the cloud
and the edges.  Besides, collaboration among users is important for training
a more robust and diverse system. Edge-cloud computing provides a
natural solution to build GenAI systems at scale. Yet, to the best of
our knowledge, there is no research addressing how a distributed system
should be designed to accommodate the computation, transmission, and exchange
of a huge amount of AIGC data. This motivates us to explore this topic
and write this overview paper. 

\begin{figure*}
\centerline{\includegraphics[width=\linewidth]{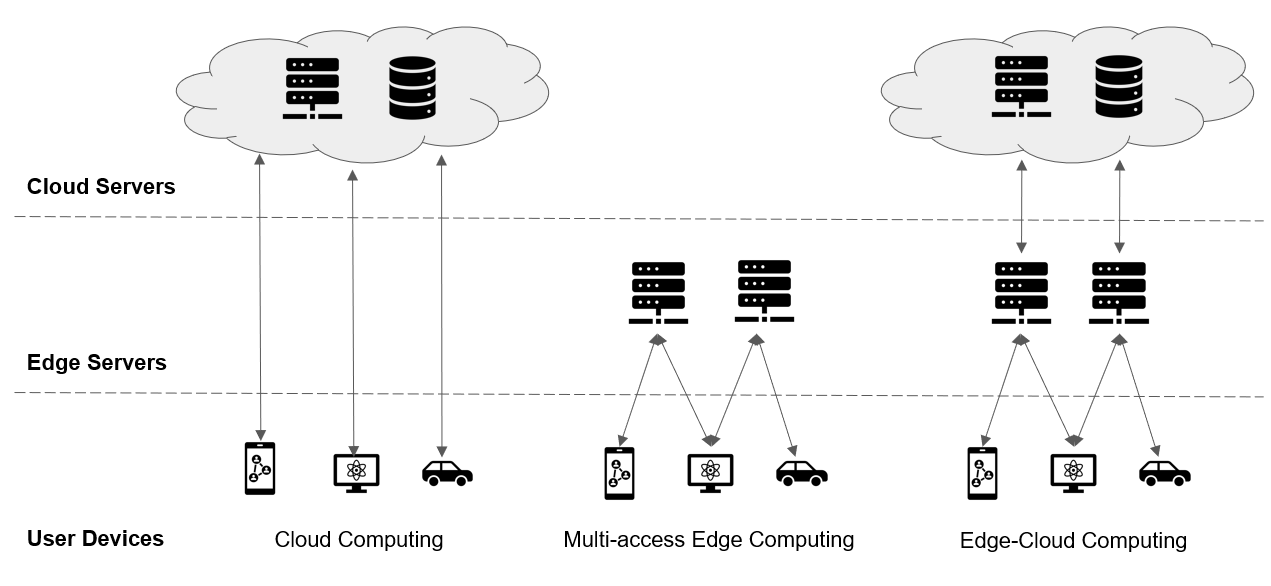}}
\caption{Three basic computing paradigms in support of large-scale computing 
systems.} \label{fig:computing}
\end{figure*}

\subsection{EDGE-CLOUD COMPUTING}

There are three basic paradigms for implementing large-scale computing
systems. They are: 1) cloud computing, 2) multi-access edge computing
(MEC), or previously mobile-edge computing, and 3) edge-cloud computing, 
as shown in Fig. \ref{fig:computing}. Among the
three, cloud computing carries out computationally demanding projects
using a large number of online servers to serve many remote users.  The
cloud has much larger computing resources than a local site.  Moving
compute-intensive tasks to the cloud has been an efficient way of data
processing.  The concept of cloud computing was introduced in the early 60s
\cite{garfinkel1999architects, surbiryala2019cloud}. It has made rapid
progress in the last several decades and become a mature business
service model. Examples include: 
Amazon Web Services (AWS)\footnote{\url{http://aws.amazon.com/ec2}}, 
Microsoft Azure\footnote{\url{http://www.microsoft.com/azure}}, 
Google Cloud Platform (GCP)\footnote{\url{https://cloud.google.com/}}, 
IBM Cloud\footnote{\url{https://www.ibm.com/cloud}}, 
Salesforce\footnote{\url{https://www.salesforce.com/}}, etc. 

As the computational power of mobile devices increases and wireless
networks become accessible at almost any place, multi-access edge computing
(MEC) provides computing, storage, and bandwidth closer to users. MEC
tends to allocate more computing tasks to the edge than the cloud.
Computation can be performed near data sources on edge devices. Edge
computing has become more important nowadays, as pointed out in a few
studies, e.g., \cite{duan2022distributed, mao2017survey, shi2016promise,
cao2020overview, elgendy2022survey}.  The MEC framework primarily relies
on edge devices, which have limited resources. In addition, the MEC framework
greatly relies on caching to improve the latency. Thus, its performance is
not good for computationally demanding tasks. 

As the demand for real-time processing, low-latency communication, and
efficient data management increases, the edge-cloud computing paradigm
emerges as a new and attractive solution. By combining the power of
cloud computing with the proximity and responsiveness of edge devices,
edge-cloud computing aims to bridge the gap between latency and
scalability.  Since it has lower latency, it is suitable for real-time
applications such as AR/VR/MR \cite{zhang2017towards, erol2018caching},
object tracking and detection \cite{ren2018distributed, tuli2019edgelens},
etc. Since it can utilize computational resources at both the cloud and edges,
it has more flexibility in load balancing to yield a more scalable solution.
Moreover, user data and privacy can be better preserved by edge-cloud
computing \cite{parikh2019security}. 

\begin{table}
\centering
\begin{tabular}{|p{40mm}|c|c|c|} \hline
Resources & Cloud Servers & Edge Servers & User Devices \\ \hline \hline
Memory & $>$24TB & $\sim$500GB & $<$64GB \\ \hline
Dist Storage & $>$25PB & $<$1PB & $<$10TB \\ \hline
Latency (RTTs) & 30 $\sim$ 50 ms & $<$10ms & - \\ \hline
Power (per year) & $>$2,000TWh & $\sim$7,500KWh & $\sim$600KWh \\ \hline
Concurrent Connections & $>$500,000 & $\sim$1,000 & 1\\ \hline
\end{tabular}
\caption{Comparison of hardware and performance specifications 
of three computational resources, namely cloud servers, edge
servers, and user devices.}\label{tab:HW_spec}
\end{table}

The hardware and performance specifications of three computational
resources (namely, cloud servers, edge servers, and user devices) are
compared in Table \ref{tab:HW_spec}. As shown in the table, cloud
servers have the highest resources in terms of computational memory and data
storage capacity.  At the same time, they have the highest power
consumption and the largest number of concurrent connections. Their
latency is also the highest since they are far from users.  It is
beneficial to shift some computation loads from cloud servers to edge
servers and user devices to balance the computational load and reduce
latency in various applications. The load-balancing idea is also called
offloading. Computation offloading \cite{feng2022computation, huda2022survey} 
and data offloading \cite{zhao2022learning, hazra2022fog} are two key
concepts in edge-cloud computing. 

\begin{figure*}[t]
\centerline{\includegraphics[width=0.8\textwidth]{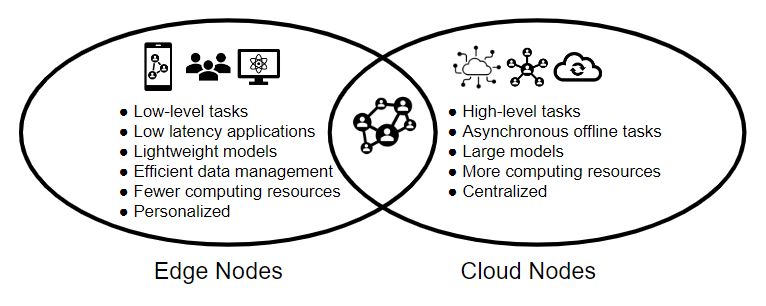}}
\caption{Roles and suitable applications for edge nodes and cloud nodes
in edge-cloud computing.} \label{fig:comp-task}
\end{figure*}

\begin{figure*}[t]
\centerline{\includegraphics[width=0.8\linewidth]{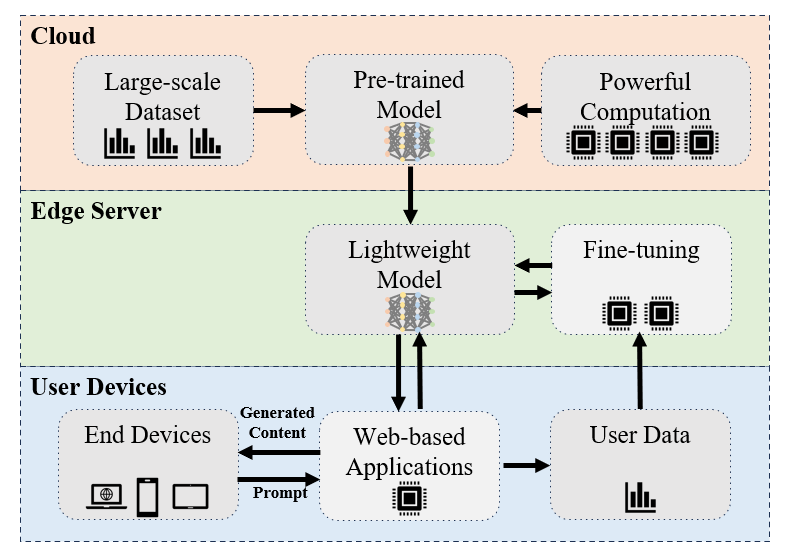}}
\caption{Implementation of GenAI systems with the edge-cloud computing paradigm.} 
\label{fig:system}
\end{figure*}

The AI tasks suitable for edge servers and cloud servers are shown in
Fig. \ref{fig:comp-task}.  Due to rich computation resources, cloud
servers can store and run large models to process high-level tasks. In
contrast, edge devices are mainly responsible for low-level
pre-processing tasks. Due to the emergence of 5G/IoT, AIGC enters a new
era.  That is, it is no longer sufficient to conduct all computations
and store all data in a centralized cloud server or data center.
Similarly, AI computation with edge servers and user devices is also not
practical in building a scalable system as AIGC data grow fast. 

Some large deep-learning AI models are difficult to deploy at the edges.
Recently, a green learning methodology \cite{kuo2022green} has been
proposed as an alternative to deep learning. Green learning AI models
have much smaller model sizes, significantly lower computational
complexity in terms of FLOPs (Floating Point Operations), faster
inference time, and less power consumption demand.
As a result,
green-learning AI models open a new door for edge servers and even user
devices in offloading cloud servers.  Hybrid deep- and green-learning
solutions match the edge-cloud computing paradigm well. That is, GenAI
has a unique mission to process low-level data and aggregate high-level
abstractions to generate creative content. GenAI can benefit the most
from the collaboration of edge and cloud servers. 

Recently, Meta announced a supercomputing cluster with very rich
computational resources\footnote{\url{https://ai.facebook.com/blog/ai-rsc/}}.
It can perform five exaflops (billion billion
calculations per second) using a total of 16,000 Nvidia A100 GPUs to
train state-of-the-art GenAI models. Servers are connected by an NVIDIA
Quantum InfiniBand fabric network with a bandwidth of 16 Tb/s to ensure
low latency in data synchronization. However, this computational scale
is not affordable for most companies and academic institutions.  Thus,
how to design scalable GenAI systems using a reasonable computing
cluster to perform similar tasks is of great interest. We put hope in
edge-cloud computing since it can leverage expandable computation
resources that are under-utilized and closer to users.

The deployment of GenAI systems on the edge-cloud computing platform is
shown in Fig.  \ref{fig:system}. Since the training of GenAI models is
most computationally heavy, it is still conducted in cloud servers.  The
training is usually done offline and asynchronous. The deployment of
trained GenAI models for the AIGC tasks can be placed as close to users
as possible to lower latency. Edge servers can be used to fine-tune
GenAI models, train personalized models, preserve user privacy, and
serve as an interface between edges and cloud servers.  It is ideal to
have several edge servers to handle individual tasks separately. More
details about the design considerations will be discussed in Sec.
\ref{sec:design}. 

\section{TWO EXEMPLARY SERVICES} \label{sec:app}

In this section, we present two exemplary GenAI services that demand low
latency and will have a large number of users when the technologies
become mature, and the markets are ready. Scalability based edge-cloud
computing is critical to their successful deployment. 

\begin{figure*}[t]
\centerline{\includegraphics[width=0.9\linewidth]{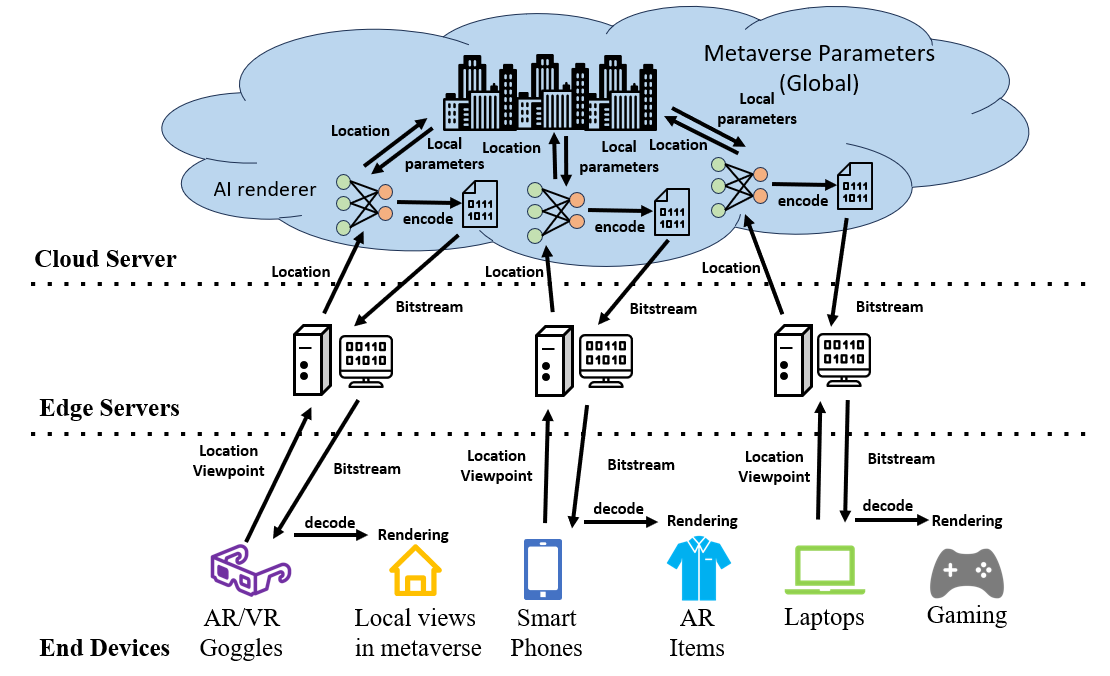}}
\caption{Illustration of the Metaverse system.} \label{fig:metaverse}
\end{figure*}

\subsection{METAVERSE SYSTEM}

Metaverse is one of the most important applications in GenAI.
With the development of GenAI, most of the generated scenes rely on
machine learning models. However, rendering is usually time-consuming.
In addition, the metaverse requires an extremely low latency in order to
make the transition more smooth and avoid dizziness when looking into
virtual reality (VR) goggles. Practically, the metaverse should also
connect every user in virtual reality. As a result, a huge map will be
required. Edge-cloud computing can be a latency-efficient solution for
future metaverse applications. For example, as illustrated in Fig.
\ref{fig:metaverse}, the entire map is stored in the centralized data
center that can be shared among all users. Then, the location, angle,
and other parameters can be collected by user devices and transmitted
through a wireless network. The cloud computing clusters are also
responsible for generating the scenes and rendering the results. The
compressed scenes will be sent back to the users. At the user end, a
lightweight decoder and renderer are deployed to display the scenes
based on the corresponding viewpoints of the users. 
As a result, such a system design can reduce the latency significantly
since the computation-heavy parts are taken care of using powerful
computation infrastructure. In addition, the amount of data transmitted
in the communication systems is minimized. The users will send the
request to the GenAI models in the cloud, and the compressed scenes will
be transmitted back to the users. 

Edge servers are a fundamental component in the metaverse system. They
serve a similar role as in the content delivery network (CDN) to distribute
content based on geographical locations and share the computation
load in the cloud server. Users in the same locations will be connected to
the same edge server. Once a user sends a request to the metaverse system to
generate the local scene, it's transmitted through the edge servers and cached.
Other users in the same location can access the cached scenes in the 
edge servers to further reduce the latency.
Computation resources in the edge servers should also be leveraged. 
For example, they can be helpful in compressing and decompressing the scenes 
generated in the cloud server. As a result, not only the latency can be 
reduced, but also the quality of the generated scenes is improved.

\begin{figure*}[t]
\centerline{\includegraphics[width=0.9\linewidth]{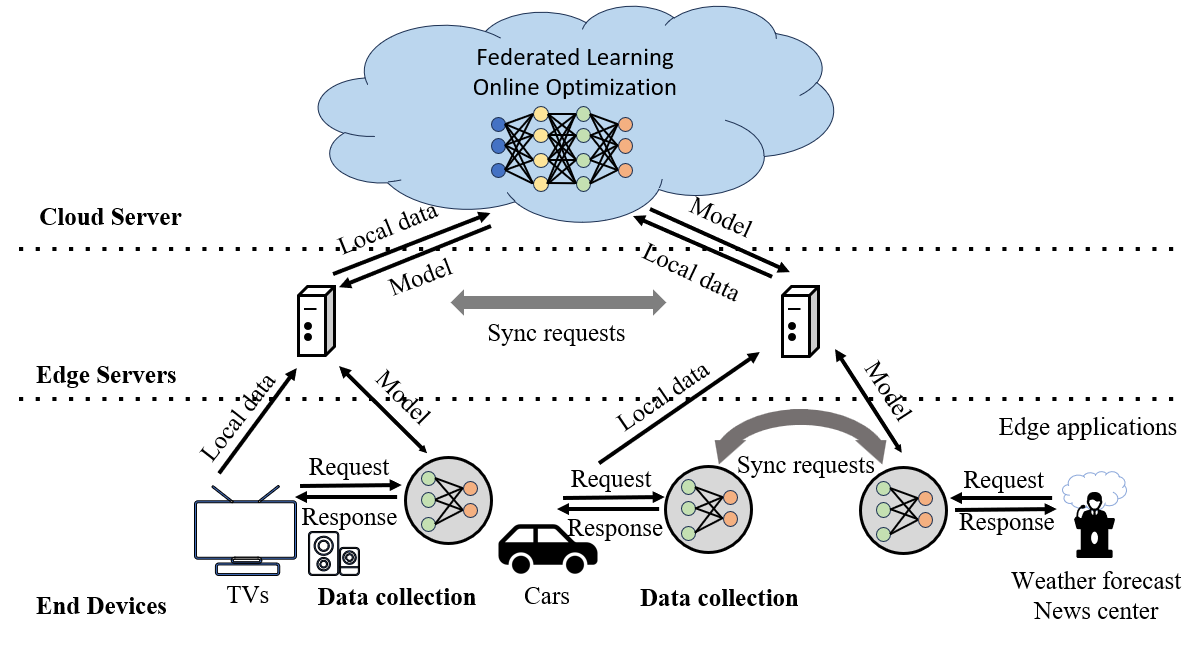}}
\caption{Illustration of the AIoT system.} \label{fig:aiot}
\end{figure*}


\subsection{ARTIFICIAL INTELLIGENCE OF THINGS}

Artificial Intelligence of Things (AIoT) is an emerging application
to combine artificial intelligence (AI) technologies in the Internet of 
Things (IoT) systems.
Through the integration of AI and the ubiquitous wireless networking
infrastructure, one can build AIoT systems, where the end
devices have certain intelligence in data processing and analytics.
GenAI can be further exploited to facilitate a broader range of
applications.  For example, a voiced assistant can interact with users
in applications such as autonomous driving, smart cities, and smart
home, where fluent human speech has to be automatically generated from
information source, which is often in the form of text data. 

To implement AIoT with edge-cloud computing, we need to take privacy,
personalization, and data synchronization into account.  Users may
collect data to train more relevant personalized GenAI models. It is
desired to train a simple GenAI model with acceptable performance on
user devices. Then, model parameters of multiple users can be sent to
cloud servers to be integrated into a more advanced GenAI model 
through federated learning \cite{nguyen2021federated, khan2021federated}.
In addition, data can be constantly collected from the end devices
to ensure the information in GenAI models is up-to-date. 
Online optimization \cite{li2019distributed, wu2019online}
supports training machine learning models with streams of data.
User devices can be synchronized with the 
advanced GenAI model through firmware updates. As a result, the whole
system can benefit from a larger pool of training data from users via
federated learning while user data privacy can still be preserved.

The hierarchy in edge-cloud computing can be utilized for more
efficient model deployment.  For example, large, middle-size, and 
lightweight models can be placed in cloud
servers, edge serves, and user devices, respectively. Different resolutions
of the models can be achieved through knowledge distillation 
\cite{wang2021knowledge, gou2021knowledge} and model parameter pruning
\cite{rui2022smart, jiang2022model}. 
The computation can be further reduced by grouping users with the same
computation facility. Different edge and cloud servers can be specialized
to process different applications efficiently. 
Personalization can be considered to optimize end devices according 
to user behavior. The personalization fine-tuning on the user devices
are generally efficient due to the deployment of lightweight models.

\section{TECHNICAL CHALLENGES}\label{sec:challenges}

There are technical challenges in the deployment of GenAI services.
The major ones include: 1) growth in model sizes, 2) power consumption, 3)
latency, and 4) infrastructure reliability. They are summarized below to demonstrate
the need for good resource coordination between edges and the cloud with
edge-cloud computing. 

\begin{figure}[t]
\centerline{\includegraphics[width=0.6\textwidth]{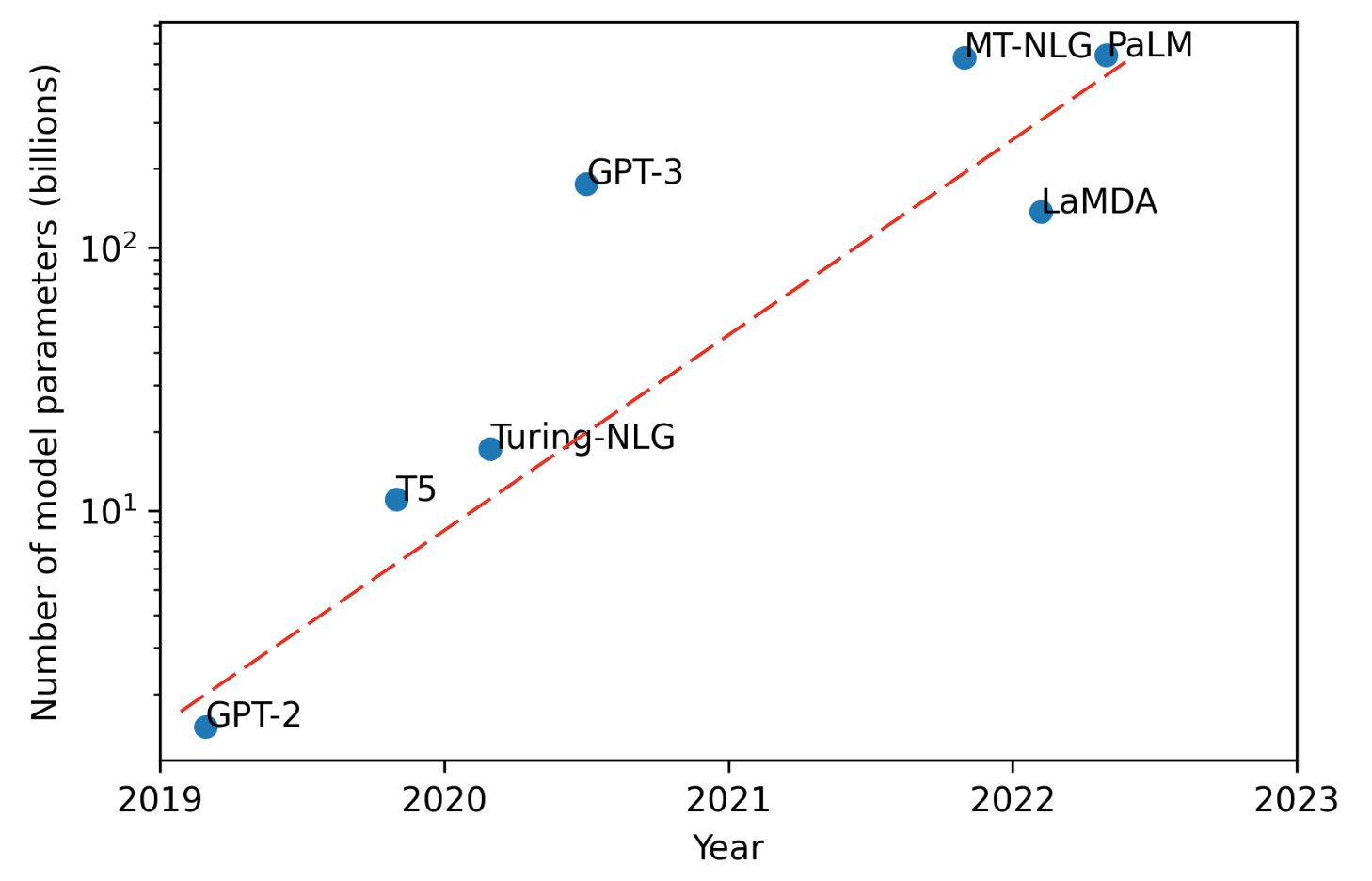}}
\caption{The development of generative LLMs and their model sizes as a function 
of time. The vertical axis is in log scale. Models in the figure include
GPT-2 \cite{radford2019language}, T5 \cite{raffel2020exploring},
Turing-NLG \cite{smith2022using}, GPT-3 \cite{brown2020language}, 
LaMDA \cite{thoppilan2022lamda}, MT-NLG \cite{smith2022using},
and PaLM \cite{chowdhery2022palm}.} \label{fig:gai-evo}
\end{figure}

\subsection{GROWTH IN MODEL SIZES}

In order to achieve better performance in various applications, GenAI
systems adopt larger models with more model parameters and computation
over time.  The growth rate of their model sizes is an exponential
function of time \cite{kaplan2020scaling} as shown in Fig.
\ref{fig:gai-evo}.  Specifically, the model sizes of neural GenAI models
double every 6 months as reported in \cite{cao2023comprehensive}. This
is called ``Moore's Law for GenAI''.  In contrast, the computation
power of CPUs and GPUs only doubles every two years in the semiconductor
manufacturing industry. If the trend continues, the demand for
computation will surpass its supply in the near future. Unless there is
a major breakthrough in supply, its limitation will hinder the future
growth of GenAI systems. Thus, how to train and run GenAI systems
through collaboration between the cloud and edges efficiently has become
an urgent issue for the entire community to tackle. 


\begin{table*}
\centering
\begin{tabular}{|l|c|c|c|c|c|c|} 
\hline
Model & Modality & Hardware & Power (watts) & Hours & Energy (kWh) & $\text{CO}_2e$ (lbs)\\ \hline \hline
WaveGAN    \cite{donahue2018adversarial} & Audio & P100 GPU x1 & 250 & 96  & 24 & 19.63\\ \hline
GANSynth   \cite{engel2019gansynth} & Audio & V100 GPU x1 & 300 & 108 & 32.4 & 26.5\\ \hline
FloWaveNet \cite{kim2018flowavenet} & Audio & V100 GPU x1 & 300 & 272 & 81.6 & 66.74 \\ \hline
BigGAN \cite{brock2018large} & Image & V100 GPU x1 & 300 & 3,072 & 921.3 & 753.54 \\ \hline
Stable Diffusion \cite{rombach2022high} & Image & V100 GPU x1 & 300 & 2,184 & 655 & 535.72 \\ \hline
GPT-2 \cite{radford2019language} & Text & TPUv3 x 32 & - & 168 & $2.8 \times 10^4$ & $2.39 \times 10^4$ \\ \hline
GPT-3 \cite{brown2020language} & Text & V100 GPU x10,000 & - & 355 & $1.29 \times 10^6$ & $1.1 \times 10^6$\\ \hline
GLaM \cite{patterson2021carbon} & Text & TPUv4s & - & - & $4.56 \times 10^5$ & $8 \times 10^4$\\ \hline
\end{tabular}
\caption{Comparison of power consumption, carbon emission, and cloud computational 
cost in the training of large GenAI models in different modalities.}
\label{tab:power}
\end{table*}

\subsection{POWER CONSUMPTION}

Power consumption is a major concern in cloud computing \cite{xu2019acg,
mirka2022generative}. The centralized computation infrastructure
consumes a significant amount of electricity in running user requests
as well as training large models.
Fig. \ref{tab:power} compares power consumption, carbon emission, and
cloud computational cost in training large GenAI models for different
modalities. The power consumption of GenAI services is even greater 
than simply training GenAI models since they need to process millions of
requests per day from the users.

Power consumption and carbon emission are closely related to
the number of floating point operations (FLOPs). More FLOPs imply higher
carbon emissions and electricity bills. 
For example, the GPT-3 model, the backbone of ChatGPT, demands $10^{23}$
FLOPs in one training iteration and $10^{15}$ FLOPs in inference.  Since
the power efficiency of CPUs/GPUs in modern computation facilities is around
$10^{10}$ FLOPs/sec-watt, it will demand 27.78 kWh ($10^5$ Joule) to process
a single request. Apparently, GenAI services are not scalable. Furthermore,
they are eco-unfriendly, unsustainable, and cost-inefficient. To achieve
sustainability with large-scale GenAI services, alternative Green
solutions under the edge-cloud computing paradigm are essential. 


\begin{figure}[t]
     \centering
  \begin{tabular}{c}
    \includegraphics[width=0.6\linewidth]{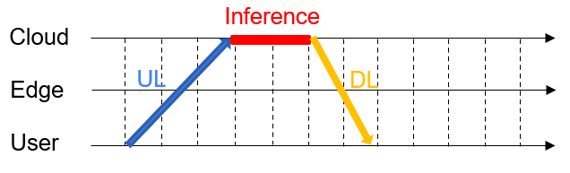}
    \\[\abovecaptionskip]
    \small (a) Cloud Computing.\\ \\
    \includegraphics[width=0.6\linewidth]{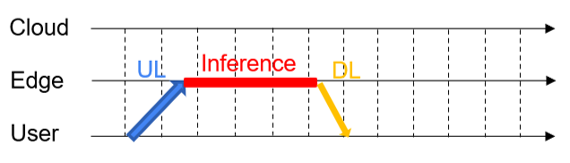}
    \\[\abovecaptionskip]
    \small (b) Multi-access Edge Computing (MEC).\\ \\
    \includegraphics[width=0.6\linewidth]{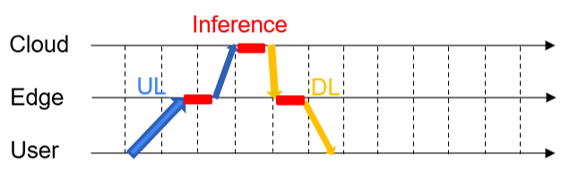}
    \\[\abovecaptionskip]
    \small (c) Edge-cloud Computing.\\ \\
  \end{tabular}
\caption{Illustration of latency in different computation frameworks.} 
\label{fig:latency}
\end{figure}

\subsection{LATENCY}

For real-time GenAI applications such
as VR and gaming, it is of uttermost importance to reduce latency. 
The latency calculation in three different computing frameworks is illustrated
in Fig. \ref{fig:latency}. It is the time between a request sent and its
response received at the user end.  It is determined by uplink
transmission time, inference time, and downlink transmission time; namely,
$$
\emph{latency} = t_{\emph{UL}} + t_{\emph{inference}} + t_{\emph{DL}},
$$
where $t_{\emph{UL}}$, $t_{\emph{inference}}$, and $t_{\emph{DL}}$
denote uplink transmission time, inference time, and downlink
transmission time, respectively. In the cloud computing framework,
the latency comes from the long uplink transmission time $t_{\emph{UL}}$ 
and downlink transmission time $t_{\emph{DL}}$.
since the computation resources are placed far from the users.
In MEC, the transmission delay is reduced since the processing units are
placed closer to the users. However, the computation resources in edge 
servers are not as powerful as the ones in the cloud servers. 
Thus, the inference time, $t_{\emph{inference}}$ will be much longer,
especially for computation-intensive applications, such as GenAI services.
In edge-cloud computing, tasks are divided efficiently between the edge and
cloud servers. Thus, the overall inference delay can be reduced by
leveraging both computation resources in edge and cloud servers. In addition,
the transmission delay is also reduced since the connection between edge servers
and the cloud is much faster than from the user ends.

For GenAI applications, their inference time can be longer than that of
other applications due to larger model sizes and more computations
required by GenAI models.  Furthermore, the output of GenAI services can
be multimedia AIGC.  Transmission of multimedia data such as video will
demand a longer downlink transmission time $t_{\emph{DL}}$ than text
data.  We can reduce $t_{\emph{DL}}$ by allocating multimedia generation
tasks to edge servers. Again, the development of green-learning-based
GenAI models are in urgent need. 

\subsection{INFRASTRUCTURE RELIABILITY}

Cloud servers need a large number of GPUs to handle user requests at
scale. As mentioned before, Meta has just started a supercomputing
center with 16,000 Nvidia A100 GPUs to support their
GenAI services. It
is unrealistic to set up such powerful but costly infrastructures in
many sites globally.  Furthermore, such a huge single-site
infrastructure is vulnerable to physical and/or cyberspace attacks.
Distributed computing with multiple lightweight cloud servers and much
more edge servers will offer a more robust AI computational
infrastructure in the future. 

\begin{figure*}[t]
\centerline{\includegraphics[width=0.8\textwidth]{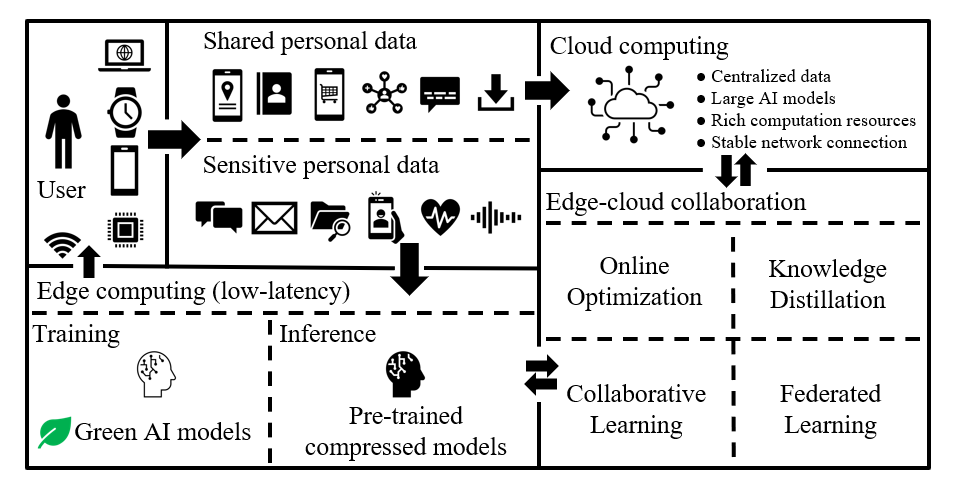}}
\caption{The roadmap of designing GenAI services at scale. 
Computation offloading, latency, privacy, and data offloading 
are the major considerations.} \label{fig:proposed}
\end{figure*}

\section{DESIGN CONSIDERATIONS} \label{sec:design}

Design considerations for providing GenAI services at scale using
edge-cloud computing are examined in this section.  Training and
deployment of GenAI services should be considered separately.  For the
training of GenAI models, a larger amount of computational resources and
training data are needed. Key considerations include: 1) computation
offloading, 2) personalization, 3) privacy, and 4) information recency.
After models are trained, it is desired to deploy them on user devices
for lower latency and power consumption. There are two main
considerations: 1) lightweight inference models, and 2) multimedia
content. For the former, lightweight models are essential because of
limited resources on edge servers and user devices. For the latter,
multimedia content will become the main media for humans to acquire
information as evidenced by the popularity of videos on the Internet
nowadays. Cross-domain content generation and interface at edges should
be considered carefully. Fig. \ref{fig:proposed} summarizes the design
considerations for providing GenAI services at scale. 

\subsection{TRAINING}

Since the training of large-scale GenAI models is costly, we need to 
consider the following issues.

\begin{figure*}[t]
\centerline{\includegraphics[width=\textwidth]{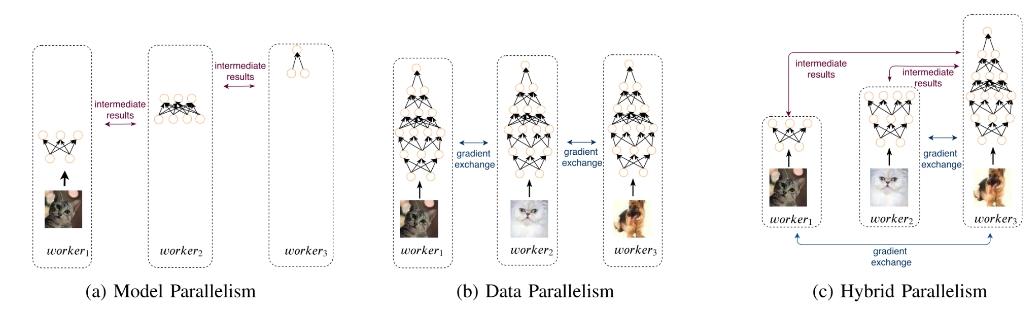}}
\caption{An existing work on computation and data offloading in DNN model training 
\cite{liu2020hiertrain}.}\label{fig:offloading}
\end{figure*}


\begin{figure}[t]
\centerline{\includegraphics[width=0.75\textwidth]{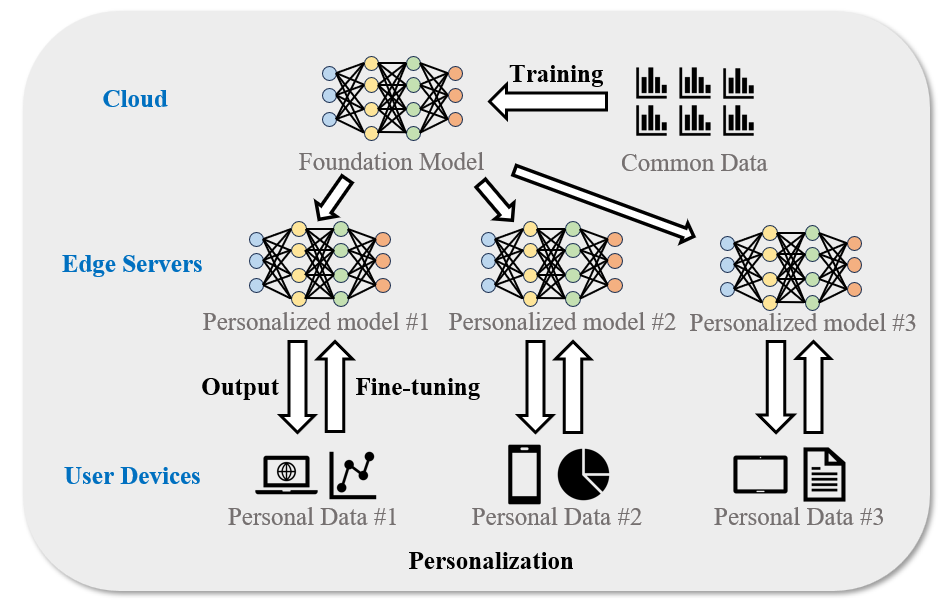}}
\caption{Personalization of GenAI services.} \label{fig:personalization}
\end{figure}

{\em a) Computation offloading.} This is an important concept in
edge-cloud computing and collaboration. It means that we need to fully
utilize computation resources in the cloud and edges. Traditional cloud
computing puts all computational loads in the cloud. Users might
experience long latency if the cloud resources cannot meet the
requirements of sudden heavy service requests. Furthermore, the
computational cost to train large GenAI models is extremely high.  It
may take days or weeks to train large models.  Thus, computation
offloading has to be considered when training GenAI systems under the
edge-cloud computing paradigm. 

Most GenAI services adopt deep neural networks (DNNs) as models.  DNNs
consist of multiple layers. To balance computation loads in training
DNNs, we can decouple the training procedure. There is an existing
work \cite{liu2020hiertrain} on DNN training under
mobile-edge-cloud computing that illustrates the concepts well, 
as shown in Fig. \ref{fig:offloading}.  
In classic cloud computing, all data are
transmitted to the cloud server, and the training takes place in the
cloud only.  In edge-cloud computing with computation offloading,
different layers can be trained by different computational facilities
(e.g. user devices, edge servers, and the cloud server). For example, as
shown in Fig. \ref{fig:offloading} (a), deeper layers are farthest from
users, and they can be trained in the cloud. Gradients are propagated to
edge servers to train middle layers.  Gradients are propagated again to
user devices.  Finally, shallow layers are closest to users, and their
parameters can be trained on user devices. As a result, system
optimization can be carried out through the collaboration of user devices,
edge servers, and the cloud server.  Only the gradient information has
to be transmitted in such a design. Another idea is to decouple the
training data as shown in Fig. \ref{fig:offloading} (b). Smaller DNNs
can be trained in parallel by leveraging data parallelism. Then,
multiple smaller models can be integrated through federated learning.
Finally, a hybrid solution exploiting both model parallelism and data
parallelism can be explored as well as shown in Fig.
\ref{fig:offloading} (c). Under the GenAI context, such parallelism and
collaboration between edges and the cloud are even more important.
Computation and data offloading should be carefully designed in
large-scale GenAI services.

{\em b) Personalization.} Edge-cloud computing can provide personalized
GenAI models.  While training a GenAI model requires a large amount of
data, personalization can be achieved by fine-tuning the trained model
with a small amount of user data. The collaboration between edges and
the cloud for personalized services is depicted in Fig.
\ref{fig:personalization}. First, an advanced GenAI model, called the
foundation model, should be trained in the cloud with common data. In
this step, the trained foundation model can handle general requests. To
achieve personalization, personal data, such as user logs and metadata,
are collected from user devices and sent to edge servers. The foundation
model is also placed in edge servers for personalization.  Then, a
fine-tuning technique can be developed to shift the model domain from a
generic one to a user-specific one using personal data.  Typically,
fine-tuning requires much fewer computation resources, and it can be
entirely conducted in edge servers. 

\begin{figure}[tb]
\centerline{\includegraphics[width=0.75\textwidth]{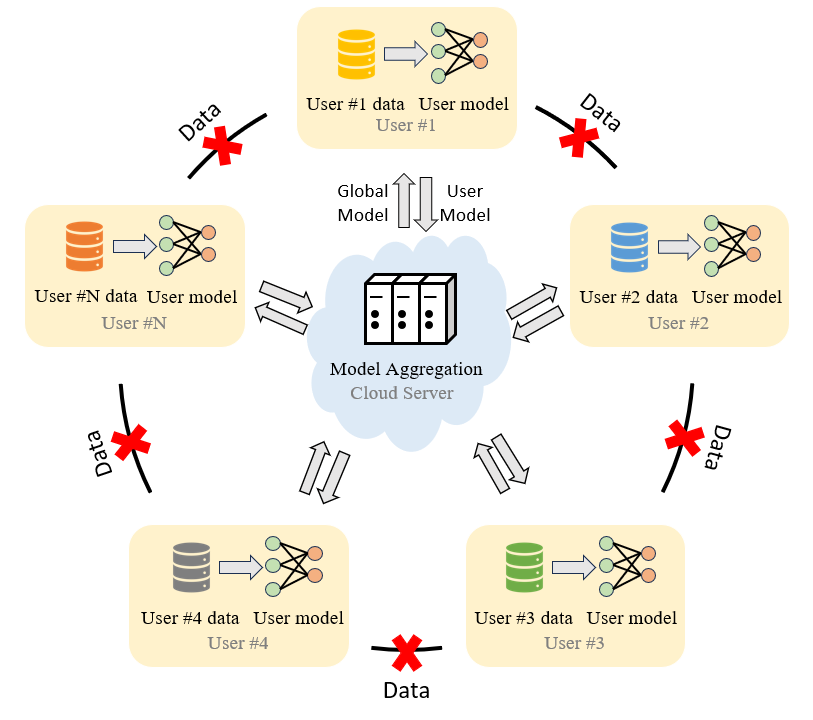}}
\caption{Privacy preservation through federated learning.} \label{fig:privacy}
\end{figure}

{\em c) Privacy.} Privacy is a major concern in GenAI services to
prevent personal information from being disclosed to other users
and companies. It is particularly important in the context of GenAI services
since generated content is difficult to control. One solution to privacy
is the use of federated learning, as shown in Fig.  \ref{fig:privacy}.
The core concept is to share the model parameters among users instead of
sharing personal data. Users will have their own models stored in user
devices or edge servers based on applications.  The models are trained
based on user data. Information exchange among users is through
aggregating user models in the cloud.  That is, all trained user models
are transmitted from edges to the cloud, where small user models are
combined to train an advanced large model.  Finally, model parameters of the
advanced model will be synchronized with user models for the next round
of training.  By sharing model parameters in federated learning, GenAI
services can preserve user privacy while collecting relevant information
from users. 


\begin{figure*}
\centerline{\includegraphics[width=0.9\textwidth]{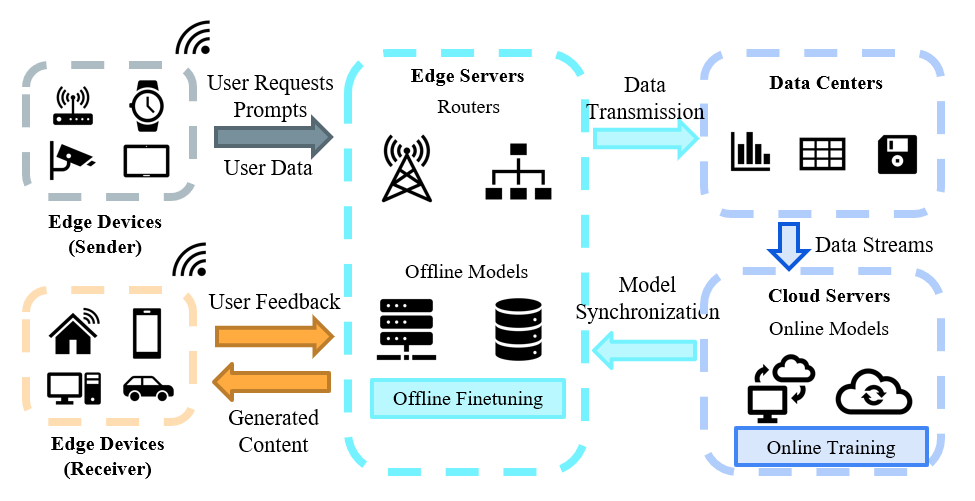}}
\caption{Online optimization in edge-cloud computing.}
 \label{fig:online}
\end{figure*}

{\em d) Information Recency.} Keeping the information updated is one of
the main challenges to GenAI services.  For example, chatbots need the
most updated information to offer a better user experience.  On the other
hand, training GenAI models is time-consuming and inefficient.
Incremental learning is needed. However, it is not easy to implement in
neural network models. Online optimization with edge-cloud computing is
an alternative way to keep the services updated. This is illustrated in
Fig.  \ref{fig:online}.  Usually, it contains two models - an online
model and an offline model.  The online model is stored in the cloud
server for the most updated information by adopting online optimization.
At the same time, a smaller offline model is placed in the edge servers
for low latency inference and cloud online model offloading.  Online and
offline models are synchronized periodically to ensure that edge
intelligence is also up-to-date. 

\subsection{DEPLOYMENT}

Two design considerations in deploying GenAI services are elaborated below.

\begin{figure*}
\centerline{\includegraphics[width=0.9\textwidth]{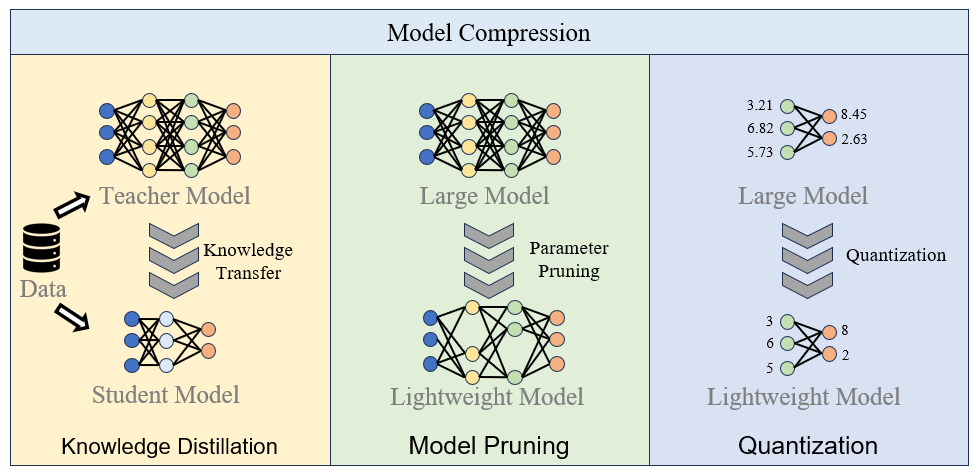}}
\caption{Existing technologies to obtain lightweight GenAI models.} \label{fig:light}
\end{figure*}

{\em a) Lightweight GenAI Models.} Deploying GenAI models on edge
servers and user devices can lower latency in user-centric applications.
Large GenAI models cannot be deployed on user devices due to their large
model sizes and high power consumption. Lightweight GenAI models, as
summarized in Fig.  \ref{fig:light} are more suitable. For example,
knowledge distillation can fit into edge-cloud computing well. With
knowledge distillation, the knowledge learned in a huge teacher model is
transferred to a smaller student model. Thus, the teacher model can be
trained and stored in the cloud server while the student model is
distilled from the teacher model in the edge servers and, then, stored
in user devices.  Model pruning adopts a similar concept to train a
smaller model from a large model, which takes place in edge servers.
Other techniques include quantization and model compression. They can
reduce the model sizes effectively without the collaboration between the
cloud and edges.

Recently, there are an increasing number of research focusing on developing lightweight
GenAI models. LLaMA \cite{touvron2023llama} reduces the number of model parameters
in LLMs to as small as 7 billion using a self-instruct training technique,
called Alpaca \cite{taori2023alpaca}.
Lightweight GenAI models encourage the development of mobile- or web-based
applications on user devices, such as WebLLM\footnote{\url{https://mlc.ai/web-llm/}}.
The small model sizes also alleviate the burden in caching-based communication
networks. Inference latency is also largely reduced due to lower computation and
transmission delay.
Research in developing lightweight GenAI models demonstrates the urgency 
to reduce the ridiculously large models while still having comparable performance.

\begin{figure*}
\centerline{\includegraphics[width=\textwidth]{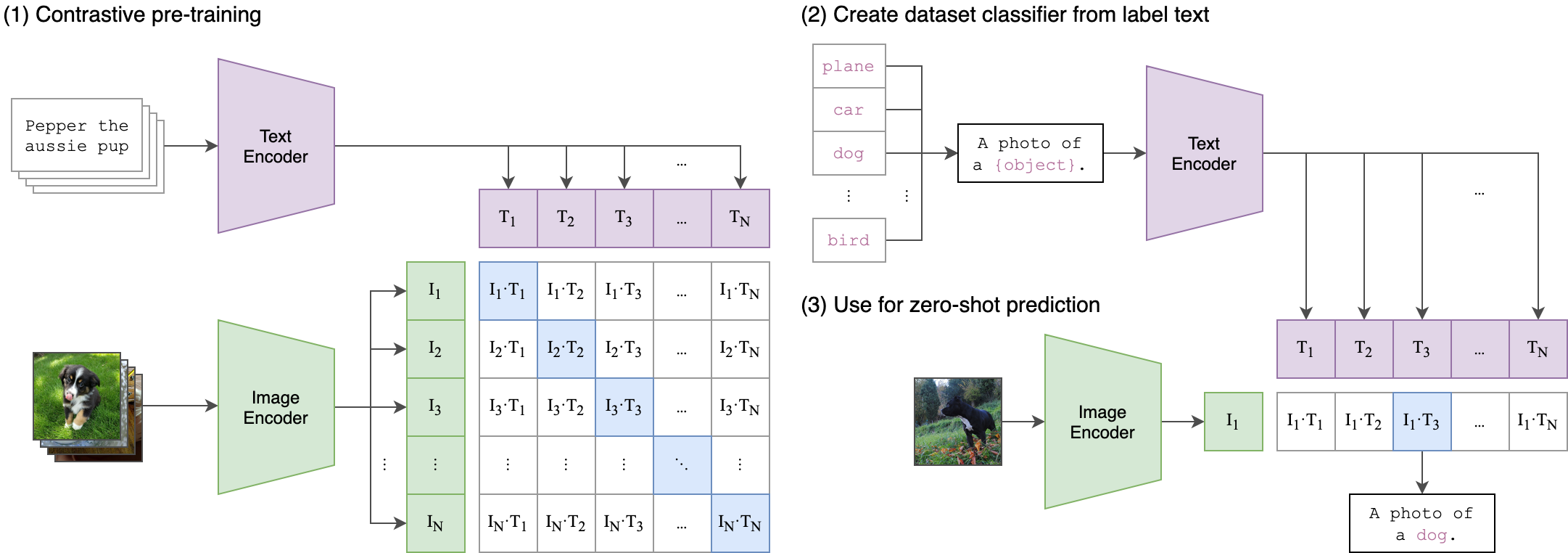}}
\caption{CLIP \cite{radford2021learning} is a text-to-image generation system that can handle two
media modalities by learning a joint embedding space.} \label{fig:multimedia}
\end{figure*}

{\em b) Cross-Domain Content Generation and Interface.} Image captioning
and text-to-image generation are two examples of cross-domain content
generation and interface. To implement cross-domain content generation,
we need a joint embedding space to connect two different modalities.
For example, as shown in Fig.  \ref{fig:multimedia}, CLIP
\cite{radford2021learning} is a well-known multi-modality GenAI model.
We elaborate on how it can be efficiently deployed under the edge-cloud
computing paradigm.  The multi-modality models usually consist of three
modules: 1) the input module, 2) the generation model, and 3) the output
module. The first and third modules are more relevant to users, and they
do not require as many computational resources as the second module.
Thus, we can place the input/output modules in edge servers or user
devices to avoid transmitting generated content.  The main generation
module is deployed in the cloud server since it requires more
computation resources. 

\section{FUTURE RESEARCH DIRECTIONS} \label{sec:future}

\subsection{GENERIC VERSUS DOMAIN-SPECIFIC GENERATIVE AI MODELS}

As one of the most famous GenAI services nowadays, ChatGPT provides a
generic GenAI model at the expense of large model size and a high
running cost. It may be advantageous to trade breadth for depth of
generated content to lower the service cost and enhance the quality of
services. For instance, the accuracy of generated content is of the top
priority in some application domains such as healthcare, financial
advice, etc.

\subsection{DECOMPOSITION OF LARGE LANGUAGE MODELS}

ChatGPT is a large generative language model built upon large pre-trained
transformers. It does not leverage the tool of knowledge graphs (KGs).
It is appealing to decompose a large language model into smaller ones
that have an interface with domain-specific KGs. This decomposition is
expected to lower the complexity of GenAI system for cost reduction.
The resulting AIGC services can be more transparent and scalable.
Furthermore, personalization is easier to offer with the help of KGs
\cite{safavi2019personalized}. That is, generic KGs are stored in the
cloud, while personalized KGs are stored in local servers or user
devices. 

\subsection{QUALITY AIGC ASSURANCE}

We have different considerations against different AIGC modalities. Two
examples are given below. 

{\em a) Visual Content.} One may use common sense to evaluate the quality
of generated visual content. For example, a picture with a person riding
a horse is more natural than the opposite. Generated content that
contradicts common sense tends to look strange to users.  Sensitive
content, copyright content, and trademarks should also be avoided in the
generated content \cite{zhang2023perceptual, du2023generative,
du2023enabling}. Automatic detection of strange and/or forbidden AIGC is
still an open problem.  Furthermore, deepfake images can be a security
concern for some applications.  A lightweight deep fake detection
solution \cite{chen2021defakehop} has been developed to address this
concern. 

{\em b) Textual Content.} The quality of generated texts can be evaluated at
three levels: grammatical correctness, readability, and factual
correctness. Coherency and conciseness are criteria for readability.
This is more difficult to evaluate than grammatical errors.
Mis/disinformation is already common over the Internet.  It will be even
easier to generate a large amount of fake news for malicious purposes
with the GenAI service. 

\subsection{GREEN GENERATIVE AI MODELS}

To address the high carbon footprint yielded by huge deep learning
networks, green learning \cite{kuo2022green} has been proposed as an
alternative learning paradigm in recent years. A green learning model is
characterized by its low carbon footprint, lightweight model, low
computational complexity, and logical transparency. Green GenAI models
have been explored in the last several years, e.g., NITES
\cite{lei2020nites}, TGHop \cite{lei2021tghop}, Pager
\cite{azizi2022pager}, GENHOP \cite{lei2022genhop}. These models are
very attractive at the edges. They can also be implemented in cloud servers
to reduce carbon footprints and save electricity bills. More efforts
along this line are needed. 

\section{CONCLUSION} \label{sec:conclusion}

The deployment of GenAI services at scale poses a new challenge to the
design of modern edge-cloud computational systems due to extremely large
model sizes, heavy power consumption, and potential latency caused by
a lack of computational and network resources.  Two illustrative GenAI
services were envisioned to show the importance of developing GenAI
systems at scale on one hand and validate the challenging claims on the
other hand in this work. Afterward, an in-depth discussion on various
design considerations of GenAI services over current communication
systems was given. It was concluded that a desired design has to balance
computational resources between edges and cloud servers and take
latency, data privacy, and personalization into account. Specifically,
federated learning is expected to play an important role, where small
GenAI models are trained at edges while large GenAI models are trained
at the cloud by combining a large number of small models. Most inference
tasks can be distributed at edges. Finally, we point out several future
research directions, such as domain-specific GenAI models, decomposition
of large language models, green GenAI models, and quality AIGC
assurance.

\bibliographystyle{unsrt}
\bibliography{references}

\end{document}